# Infrared Optical Anisotropy in Quasi-1D Hexagonal Chalcogenide BaTiSe$_3$

*Boyang Zhao, Hongyan Mei, Zhengyu Du, Shantanu Singh, Tieyan Chang, Jiaheng Li, Nicholas S. Settineri, Simon J. Teat, Yu-Sheng Chen, Stephen B. Cronin, Mikhail A. Kats, and Jayakanth Ravichandran*

Boyang Zhao, Zhengyu Du, Shantanu Singh, Jayakanth Ravichandran
Mork Family Department of Chemical Engineering and Materials Science, University of Southern California, Los Angeles, California, 90089, USA
E-mail: j.ravichandran@usc.edu

Hongyan Mei, Mikhail A. Kats
Department of Electrical and Computer Engineering, University of Wisconsin–Madison, Madison, WI, 53706, USA.

Tieyan Chang, Yu-Sheng Chen
NSF's ChemMatCARS, The University of Chicago, Lemont, IL, 60439, USA.

Jiaheng Li
Beijing National Laboratory for Condensed Matter Physics and Institute of Physics,
Chinese Academy of Sciences, Beijing 100190, China

Nicholas S. Settineri, Simon J. Teat
Advanced Light Source, Lawrence Berkeley National Laboratory, Berkeley, California 94720, USA

Stephen B. Cronin, Jayakanth Ravichandran
Ming Hsieh Department of Electrical Engineering, University of Southern California, Los Angeles, California, 90089, USA

Jayakanth Ravichandran
Core Center of Excellence in Nano Imaging, University of Southern California, Los Angeles, California, 90089, USA

**Abstract**

Polarimetric infrared detection bolsters IR thermography by leveraging the polarization of light. Optical anisotropy, *i.e.,* birefringence and dichroism, can be leveraged to achieve polarimetric detection. Recently, giant optical anisotropy was discovered in quasi-1D narrow-bandgap hexagonal perovskite sulfides, $A_{1+x}TiS_3$, specifically $BaTiS_3$[1,2] and $Sr_{9/8}TiS_3$[3,4]. In these materials, the critical role of atomic-scale structure modulations[4,5] in the unconventional electrical[5,6], optical[7,8], and thermal[7,9] properties raises the broader question of other materials that belong to this family. To address this issue, for the first time, we synthesized high-quality single crystals of a largely unexplored member of the $A_{1+x}TiX_3$ (X = S, Se) family, $BaTiSe_3$. Single-crystal X-ray diffraction determined the room-temperature structure with the *P*31*c* space group, which is a superstructure of the earlier reported[10] *P*6$_3$/*mmc* structure. The crystal structure of $BaTiSe_3$ features antiparallel *c*-axis displacements similar to $BaTiS_3$,[2] but is of lower symmetry. Polarization-resolved Raman and Fourier transform infrared (FTIR) spectroscopy were used to characterize the optical anisotropy of $BaTiSe_3$, whose refractive index along the ordinary (⊥ *c*) and extraordinary (∥ *c*) optical axes was quantitatively determined by combining ellipsometry studies with FTIR. With a giant birefringence $\Delta n$~0.9, $BaTiSe_3$ emerges as a new candidate for miniaturized birefringent optics for mid-wave infrared to long-wave infrared imaging.

# 1. Introduction

Mid-wave infrared (MWIR) and long-wave infrared (LWIR) imaging, widely used in thermographic systems for remote sensing[11], free-space telecommunication[12], and surveillance[13], utilizes the MWIR (~3-5 μm) and LWIR (~8-14 μm) atmospheric transmission windows[14] to see beyond human vision. On the other hand, due to the intrinsic and environmental radiation noise[15], thermography has limited temporal and spatial resolution[16] despite being in high demand for clinical diagnostics[17] and face recognition[18]. Polarimetric IR imaging, performed often using birefringent optics, improves the contrast of thermography and minimizes imaging artifacts. Since most natural reflections and artificial radiations are partially polarized, polarimetry can significantly improve thermography[19]. However, conventional birefringent crystals such as $TiO_2$[20], $CaCO_3$[21], $LiNbO_3$[22], and $YVO_4$[23] often have appreciable absorption for MWIR and LWIR, and liquid crystals[24] are commonly opaque in the MWIR and LWIR due to resonant IR absorptions[25] of light elements. This calls for the development of materials with broadband transparency and appreciable optical anisotropy in the MWIR and LWIR.

Hexagonal sulfides $A_{1+x}TiS_3$ (A = Ba and Sr) with quasi-one-dimensional (quasi-1D) chains of faced-shared $TiS_6$ octahedral[1] have giant optical anisotropy between the inter-chain (*a-b* plane) and intra-chain (*c*-axis) orientations[3,7] across the MWIR to LWIR regions. We recently found that the optical anisotropy of crystalline $A_{1+x}TiS_3$ can originate from not only its anisotropic crystal structure but also due to structural modulations in *non-stoichiometric* compounds[4] or due to correlated disorder of the $TiS_6$-octahedra[2]. In this work, we extend the class of quasi-1D birefringent chalcogenide crystals $A_{1+x}TiX_3$ (A = Ba, Sr; X = S, Se) chalcogenides by demonstrating comparable optical anisotropy in single crystals of $BaTiSe_3$, grown for the first time. At the same time, we studied the structural similarities and differences compared to $BaTiS_3$ by carefully mapping the structure of $BaTiSe_3$ using single-crystal X-ray diffraction and Raman spectroscopy. The larger real part of the refractive index ($n_o$ ~ 3.2, $n_e$ ~ 4.1), compared to $BaTiS_3$ ($n_o$ ~ 2.61, $n_e$ ~ 3.37), with a greater birefringence of $\Delta n$ ~ 0.9 in $BaTiSe_3$ enables additional design flexibility for birefringent optics operating in the MWIR to LWIR.

# 2. BaTiSe$_3$ Crystal Growth

$BaTiSe_3$ crystals were synthesized using chemical vapor transport with iodine as the transport agent in a sealed quartz ampoule. The synthesis process is similar to that used for $BaTiS_3$[7] and $Sr_{9/8}TiS_3$[3], with high-purity selenides in place of sulfide precursors (see more details in Methods and Supporting Information). The synthesis temperature is around 945 °C to complete the reaction while maintaining a steady crystal growth with the help of an iodine transport agent. BaTiSe$_3$ crystals preferably crystallize on top of the quartz ampule wall and grow along it (Figure S1a), giving rise to a thin needle-like (Figure 1a and Figure S1c) or thicker rod-like (Figure S1e) geometrical shapes, the latter may achieve their size with much greater concentration twin domain boundaries (or greater crystal mosaicity). Chemical composition mapping *via* energy dispersion spectroscopy (EDS) attached to a scanning electron microscope (SEM) reveals uniform chemical composition across the crystal (Figure 1a), with an overall integrated atomic ratio Ba:Ti:Se around 2.1: 1.9: 6.0 (Figure 1b). This is close to the nominal stoichiometry of 1:1:3 well within the accuracy limits of EDS. Moreover, over the 1:1:3 stoichiometry is also validated by the X-ray diffraction studies, which will be discussed later.

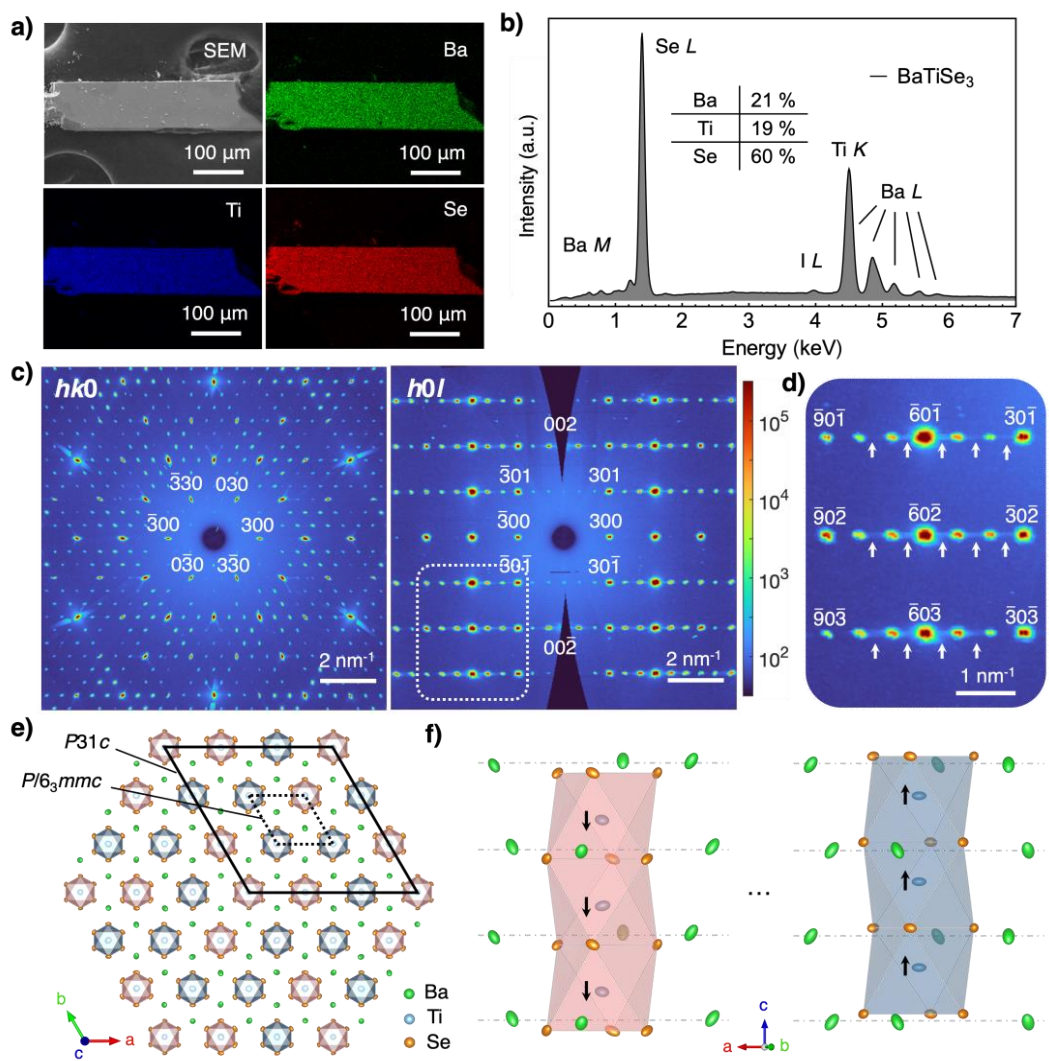

**Figure 1.** Characterization of BaTiSe$_3$ single crystals. **(a)** Scanning electron microscope (SEM) image and energy dispersion spectroscopy (EDS) mapping of BaTiSe$_3$ crystal. Ba, Ti, and Se

are uniformly distributed across the crystal **(b)** Quantitative chemical composition analysis of EDS spectrum. The atomic ratio of Ba:Ti:Se is around 0.21:0.19:0.60, close to the theoretical 1:1:3. **(c)** X-ray diffraction studies of BaTiSe$_3$ crystals show a superstructure of the formerly reported[10] *P6$_3$/mmc* along the *a-b* plane, represented by (left) *hk*0, and (right) *h0l* or *0kl* precession reciprocal maps from single-crystal X-ray diffraction (SC-XRD). As we look closer at the intensity distribution of observed reflections, such as within the dashed box, reflections in **(d)** show streak-like diffuse scattering within the *a-b* plane (pointed out by arrows) but are sharp along the *c*-axis. Such is a sign of correlated *a-b* plane disorder. **(d)** BaTiSe$_3$ is then refined as a *P31c* space group in a $2\sqrt{3} \times 2\sqrt{3} \times 1$ supercell of the *P6$_3$/mmc* along the *a-b* plane. **(e)** In *P31c*, TiS$_6$ chains are displaced along the *c*-axis in opposite directions, colored in red (downwards along the *c*-axis) and blue (upwards along the *c*-axis) as indicated in **(d)** and **(e)**. Besides, Ti atoms also move away from the centroid of Se$_6$ octahedra along the *c*-axis. This is analogous but of different periodicity compared to BaTiS$_3$[2].

Crystal orientation is determined by out-of-plane X-ray diffraction (XRD) while varying the tilting angle ($\chi$) about the rotational axis parallel to the long edge (Figure S2b) of the needle, shown in Figure S2a. Observed XRD peaks match the reported *P6$_3$/mmc* structure[10] and are indexed based on the 2$\theta$ position. The rocking curve of 020 has a full width at half maximum (FWHM) of 0.07°, showing great crystallinity of needle-like BaTiSe$_3$. Upon tilting about the long-axis of the needle-like BaTiSe$_3$, the 100-series reflections are observed around 60° tilt from each other, while 210 and 120 are observed ~ 30° tilt about 100 or 010. We thus confirmed the long axis of the BaTiSe$_3$ needle as the rotational axis (the *c*-axis). However, as we look closer at the off-principle-axes (not along 100, 010, or 001) reflections, the room-temperature structure of BaTiSe$_3$ reveals periodic symmetry breaking similar to the supercell modulation in BaTiS$_3$[2,5].

## 3. In-depth Crystal Structure Determination

Single-crystal X-ray diffraction (SC-XRD) resolves the crystal structure by mapping the reciprocal space diffraction patterns to atomic-scale resolution[26]. SC-XRD was carried out on BaTiSe$_3$ crystals at room temperature (Table I) with high-flux X-ray synchrotron radiation at Lawrence Berkeley National Laboratory and Argonne National Laboratory. Figure 1c shows the precession maps extracted from SC-XRD of BaTiSe$_3$ along *h0l* and *hk*0 reciprocal planes. Although major reflections are roughly consistent with the formerly reported[10] *P6$_3$/mmc*, symmetric weak reflections are observed, whose periodicity matches exactly with a $2\sqrt{3} \times 2\sqrt{3} \times 1$ superstructure.

We then refine[27] BaTiSe$_3$ in 21.1 Å × 21.1 Å × 6.1 Å unit cell as a *P*31*c* space group (Table S1 lists the refinement statistics including a refinement residual $R_1$ as low as 2.66%). The *P*31*c* BaTiSe$_3$ breaks the translational, inversion, and mirror symmetries by adopting ordered *c*-axis off-centric TiSe$_6$ octahedral in *c*-axis antiparallel displaced TiSe$_6$-chains, which was also observed in the room temperature crystal structure of BaTiS$_3$[2]. Figure 1e visualizes the resulting (Table S2 and S3) supercell-ordering of the antiparallel TiS$_6$ chain displacements, whose *P*31*c* unit cell is highlighted (solid line) in comparison with *P*6$_3$/*mmc* (dotted line). TiSe$_6$ chains are colored red (-*c*) and blue (+*c*) corresponding to the direction of chain displacement to the Ba atoms around them along the *c*-axis, illustrated in Figure 1f. On top of the chain displacements, Ti atoms also displace away from the centroid of TiSe$_6$ octahedra, thus leaving behind antiparallel TiSe$_6$ dipoles along the *c*-axis.

Moreover, as we examine the scattering profile between the Bragg reflections, diffuse scattering arises. Figure 1d points out the streaky diffuse scattering between adjacent sharp Bragg peaks, observed only along the *a-b* plane. This is analogous to the *a-b* plane diffuse scattering in BaTiS$_3$[2]. Consequently, the corresponding electron density distribution of Ti atoms along the *a-b* plane adopts the anharmonic trigonal displacements (Figure S3) in the same manner as the correlated *a-b* plane displacements of room temperature BaTiS$_3$, which has been shown as a significant factor in achieving the giant optical anisotropy[2] of BaTiS$_3$. We thus expect analogous disordered Ti *a-b* plane displacements[2] in BaTiSe$_3$ to play an important role in achieving optical anisotropy, comparable to BaTiS$_3$, across the visible to infrared wavelength range.

## 4. BaTiSe$_3$ Optical Anisotropy

In line with the lower symmetry room-temperature structure of BaTiSe$_3$ (21.1 Å × 21.1 Å × 6.1 Å unit cell of *P*31*c* space group) than BaTiS$_3$ (13.3 Å × 13.3 Å × 5.8 Å unit cell of *P*6$_3$*cm* space group[2]), more active vibrational Raman modes[28] compared to BaTiS$_3$ were observed in the Raman spectra as shown in Figure 2a. Under 532 nm excitation, the Raman spectrum resolves at least six Raman modes: 121 cm$^{-1}$, 158 cm$^{-1}$, 186 cm$^{-1}$, 227 cm$^{-1}$, 252 cm$^{-1}$, 315 cm$^{-1}$, marked by dashed lines. This is a clear distinction from two major Raman modes observed in BaTiS$_3$ under the same conditions[29].

We studied the polarization-dependent Raman response by varying the polarization of the 532 nm excitation, as shown in Figure 2b. We represent the polarization dependence of Raman mode intensity (before background subtraction) as a polar plot in Figure 2c to show the anisotropic Raman response. With the *a*-axis aligned parallel to the 0° polarization, 121 cm$^{-1}$ and 315 cm$^{-1}$ are strongest near 0°, while 186 cm$^{-1}$ and 227 cm$^{-1}$ (along with difficult-to-extract 158 cm$^{-1}$ and 252 cm$^{-1}$) near 90°. Although a polarization analyzer is needed to map the exact Raman tensors of observed vibrational modes, it is evident that qualitatively BaTiSe$_3$ shows a highly anisotropic Raman response for 532 nm excitation.

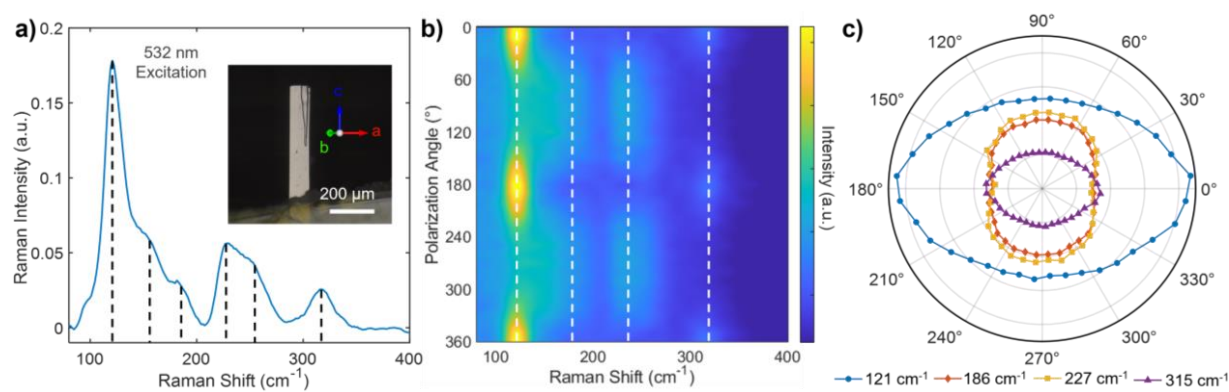

**Figure 2.** Room-temperature polarization-resolved Raman study of BaTiSe$_3$. **(a)** Unpolarized Raman spectrum of BaTiSe$_3$. Major Raman modes are labeled by dashed lines. The inset shows the optical image of the BaTiSe$_3$ crystal whose crystallographic orientation is labeled. **(b)** Polarization-resolved Raman spectrum with linearly polarized excitation light 0° to 360° with respect to the *a*-axis. **(c)** Area-under-the-curve Raman intensities of different polarization orientations. 121 cm$^{-1}$ and 315 cm$^{-1}$ show opposite polarization dependence compared to 186 cm$^{-1}$ and 227 cm$^{-1}$.

Infrared (IR) reflectance and transmittance spectra were measured by polarization-resolved Fourier transform infrared spectroscopy (FTIR). When incident light was linearly polarized parallel or perpendicular to the uniaxial optical axis, which is the *c*-axis based on the trigonal *P*31*c* crystal structure, anisotropic transmittance [Figure 3a] and reflectance [Figure 3b] spectra were observed. The limited thickness uniformity, as shown in the inset of Figure 3a, and the crystal tilting perturbs the Fabry-Perot fringes in the transmittance spectra and leads to appreciable scattering. Nevertheless, the anisotropic absorption edges of the extraordinary (∥ *c*) and ordinary (⊥ *c*) optical polarization were clear and extracted to be 0.20 eV and 0.33 eV.

Further, the Fabry-Perot fringes of the reflectance spectra (Figure 3b), measured on the smooth front surface, show extinction at the energies consistent with the absorption edges seen in the transmission spectra [Figure 3a].

As the refractive index is relatively constant within the transparent regime, the Fabry–Pérot interference between the top and bottom surfaces of BaTiSe$_3$ has a free spectral range[30] of:

$$\Delta \nu \approx \frac{c}{2nl \cos\theta} = \frac{c}{2l \cos\theta} \cdot \frac{1}{n}$$

, where $\nu$ is the frequency, $c$ is the vacuum velocity of light, $n$ is the real part of the refractive index, $l$ is the crystal thickness ~21.33 µm, and $\theta$ is the incident angle. We thus carry out a fast Fourier transform (FFT) for the Fabry–Pérot fringes on the reflectance spectrum on the significantly smoother top surface [Figure 3b] to approximate the spectrum-averaged real part of the refractive index ($n$)[31] between the 0.075 eV and 0.2 eV. The inset of Figure 3b shows the resultant FFT spectrum of refractive index, with a birefringence of $n_e - n_o$ ~ 0.9.

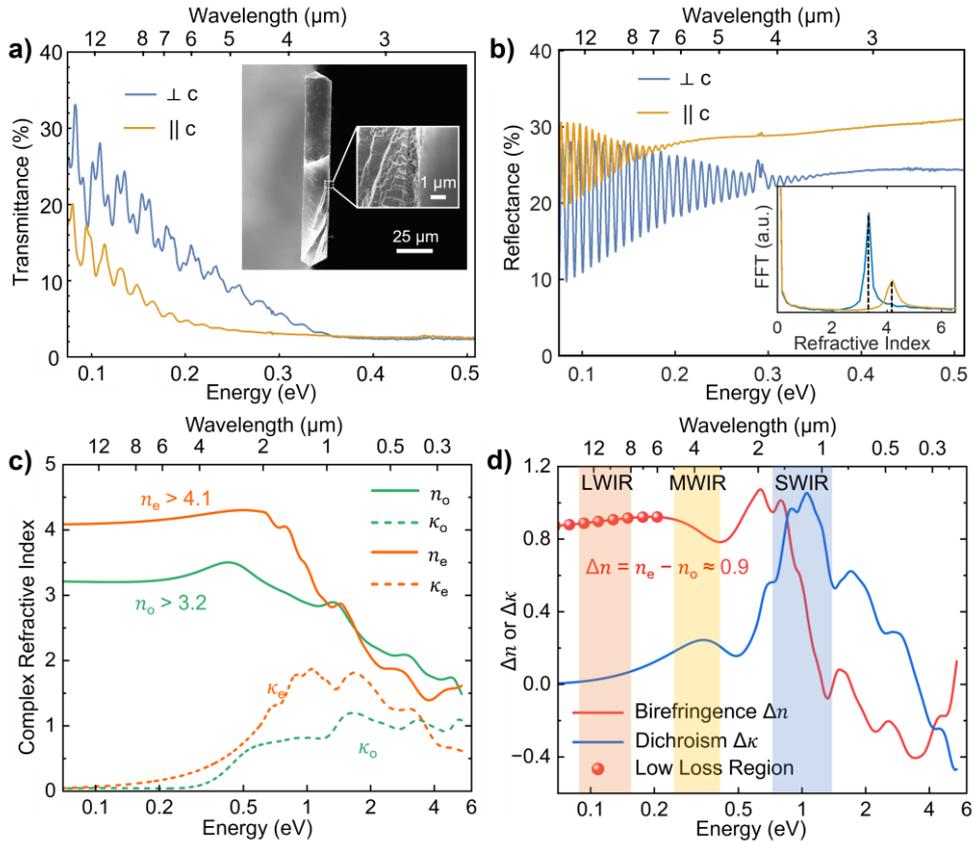

**Figure 3.** Optical anisotropy of BaTiSe$_3$. Polarization-resolved FTIR **(a)** transmittance and **(b)** reflectance spectra show large anisotropy between the ordinary (⊥ c) and the extraordinary (|| c) polarization. The absorption energy of ordinary and extraordinary are observed at 0.20 eV

and 0.33 eV, consistent with the extinction of Fabry-Perot fringes. The cross-section of BaTiSe$_3$ is shown in the inset of (**a**). The crystal thickness was measured to be 21.33 μm. Inset of (**b**) shows FFT analysis on the reflectance spectra for the polarization-dependent refractive index of BaTiSe$_3$. (**c**) Optical properties obtained from combining FTIR and ellipsometry, both real ($n$) and imaginary ($\kappa$) parts for ordinary and extraordinary linear polarization. (**d**) The large dichroism ($\Delta\kappa = \kappa_o - \kappa_e$) peak across SWIR leaves behind a giant birefringence ($\Delta n = n_e - n_o$) up to 0.9 across the MWIR to LWIR wavelengths.

To fully quantify the degree of optical anisotropy, we then combined variable-angle ellipsometry measurements over the spectral range of 210 nm to 2500 nm with the polarization-resolved reflection and transmission measurements over 1.5 μm to 17 μm and quantitatively extracted the complex refractive index of BaTiSe$_3$ for wavelengths from 210 nm (5.9 eV) to 17 μm (0.073 eV) in Figure 3c (*see more details in Supporting Information Section V*). The resulting birefringence ($\Delta n = n_e - n_o$) and dichroism ($\Delta\kappa = \kappa_o - \kappa_e$) are then plotted in Figure 3d. The dichroism becomes largest near 1 eV, within the short-wave infrared (SWIR) spectrum range, and the birefringence is as large as 0.9 across MWIR and LWIR. As the low-loss regime of BaTiSe$_3$ occurs below 0.20 eV, red spheres marked its spectrum range in Fig. 3d.

First-principles calculations (for details see Methods and Figure S8) were carried out on the *P*31*c*-BaTiSe$_3$, without considering the disordered *a-b* plane Ti displacements. Figure S9 compares the experimental and calculated birefringence of BaTiSe$_3$. Since *P*31*c* represents only the ordered structural characteristics of BaTiSe$_3$, the calculated birefringence roughly captures the magnitude and trend of the birefringence spectrum but does not represent the entire birefringence spectrum, especially below 0.5 eV. A similar near-Fermi level optical structure mismatch is corrected by introducing the disordered *a-b* plane Ti displacements in BaTiS$_3$[2]. We then presume the anomalous but more complicated BaTiSe$_3$ *a-b* plane Ti displacements are one of the leading factors towards the birefringence of BaTiSe$_3$.

We, therefore, compare the optical anisotropy of A$_{1+x}$TiX$_3$ (A = Sr, Ba; X = S, Se). The imaginary part of the refractive index (or extinction coefficient), $\kappa_o$ and $\kappa_e$, of BaTiS$_3$[1], BaTiSe$_3$, and Sr$_{9/8}$TiS$_3$[4] are compared with each other in Figure 4a. Contrary to the $\kappa$ peak featuring a sharp drop of absorption that decays to ~ 0 near 4 μm in in Sr$_{1+x}$TiS$_3$, BaTiS$_3$ and BaTiSe$_3$ have $\kappa$ peaks at shorter wavelengths but decay gradually to zero until ~5 μm and ~ 6 μm. Such absorption characteristic of BaTiS$_3$ and BaTiSe$_3$ leads to dichroism of a broader spectrum range but also cause more loss (reflectance + transmittance < 100%) in the LWIR region compared to $Sr_{9/8}TiS_3$.

The room temperature birefringence ($\Delta n$) and the low loss regions (after the absorption edge) of $BaTiS_3$[1], $BaTiSe_3$, and $Sr_{9/8}TiS_3$[4] are then listed for wavelengths from 1 µm to 17 µm in Fig. 4, along with a variety of IR birefringent crystals[20–23,32–36]. $BaTiS_3$ and $BaTiSe_3$ share similar structural features, especially the existence of the *a-b* plane Ti displacements and the disorder of such in-plane distortions. $BaTiSe_3$, up to 0.9, slightly surpasses the birefringence of $BaTiS_3$, which is up to 0.76. $Sr_{9/8}TiS_3$ and other $Sr_{1+x}TiS_3$, however, distort $TiS_6$ chains in an incommensurate manner, periodically introducing rotational distortion to $TiS_6$ polyhedron. The presence of localized Ti *d*-states, commensurate with the excess Sr, significantly increases the birefringence of $Sr_{9/8}TiS_3$ to a much higher region, up to 2.1.[4]

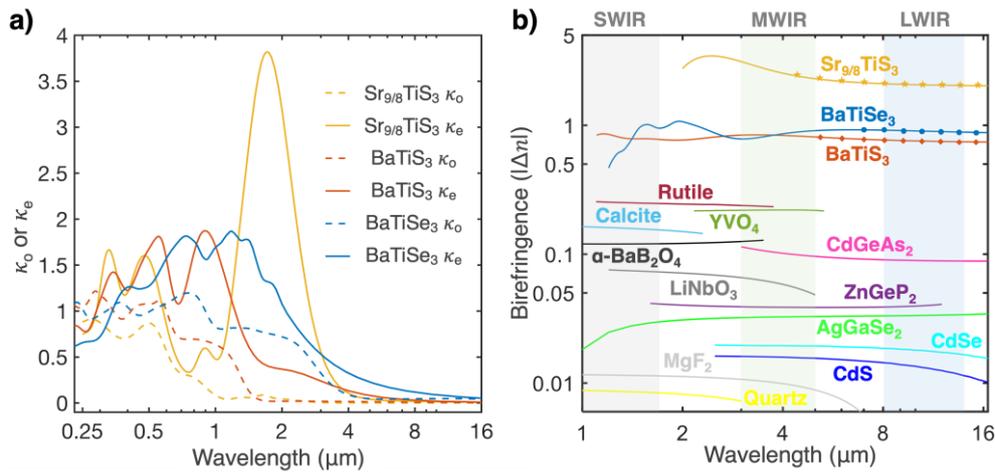

**Figure 4.** Comparison of the refractive index of $A_{1+x}TiX_3$ (A = Sr, Ba; X = S, Se). **(a)** Extinction coefficients, $\kappa_o$ and $\kappa_e$, of $Sr_{9/8}TiS_3$, $BaTiS_3$ and $BaTiSe_3$. **(b)** The absolute birefringence value of a variety of IR birefringent crystals (from the literature[1,4,20–23,32–36]), quasi-1D perovskite chalcogenides $Sr_{9/8}TiS_3$, $BaTiS_3$, and $BaTiSe_3$ extends the largest birefringence to ~0.7 - 2.1. The symbols indicate low-loss regions of $BaTiSe_3$, $BaTiS_3$, and $Sr_{9/8}TiS_3$.

## 5. Conclusions

In this article, we demonstrated another member of the quasi-1D perovskite chalcogenide $A_{1+x}TiX_3$ (A = Sr, Ba; X = S, Se), $BaTiSe_3$, to be highly optically anisotropic at room temperature. We synthesized and studied the single crystals of $BaTiSe_3$, formerly reported[10] $P6_3/mmc$ space group in powder form. High-quality $BaTiSe_3$ crystals adopt antiparallel displacements and dipolar octahedron symmetry breaking along the *c*-axis whose periodicity follows a 2√3×2√3×1 supercell of *P*31*c* symmetry. Despite the longer-range *a-b* plane structural complexity compared to BaTiS$_3$, BaTiSe$_3$ crystals possess an analogous uniaxial optical axis parallel to the long axis of the needle, with giant birefringence (Δ*n* up to 0.9) in a broadband spectral region across the MWIR and LWIR ranges. We anticipate the class of quasi-1D chalcogenides, ABX$_3$ (A = Sr, Ba, ..., B = Ti, Zr, …, and X = S, Se, …) to cover a wide spectral range of optical and optoelectronic anisotropy, leading to miniaturized polarization resolved IR devices for sensing and telecommunications.

**METHODS**

*Crystal Growth*: Single crystals of BaTiSe$_3$ were grown by chemical vapor transport with iodine as a transporting agent. Stoichiometric amounts of barium selenide, titanium, and selenium were mixed and loaded in a nitrogen-filled glovebox with iodine in a quartz ampule before being evacuated and sealed with a blowtorch. The sealed ampoule was heated to the reaction temperature of 945 °C at 100 °C/h and dwelled for 100 h before being cooled down within the tube furnace. More details in *Supporting Information Section I*.

*X-ray Diffraction*: XRD studies were performed in a Bruker D8 Advance X-ray diffractometer (XRD) in parallel beam configuration, using a germanium (004) two-bounce monochromator for Cu *Kα*1 ($\lambda$ = 1.5406 Å) radiation. Crystals are loaded on the Compact Cradle on top of a glass-slide holder and aligned. Diffractions were carried out in the out-of-plane mode while sweeping the χ-tilting. The rocking curve was then measured for the strongest 200 reflections.

*Synchrotron Single-Crystal X-ray Diffraction*: Synchrotron single-crystal X-ray diffractions were carried out at beamline 12.2.1 at the Advanced Light Source (ALS), Lawrence Berkeley National Laboratory and beamline 15 ID-B at the Advanced Photon Source, Argonne National Laboratory. Crystals with pre-determined chemical composition were submerged in oil and cut to the preferred sizes before mounting to the MiTeGen Kapton loops for SC-XRD.

At ALS 12.2.1, the crystal was placed in a nitrogen cold stream on the goniometer head of a Bruker D8 diffractometer, which is equipped with a PHOTON100 CMOS detector operating in shutter-less mode. Diffraction data were collected using synchrotron radiation monochromate with a wavelength of 0.72880 Å with silicon (111).

At APS 15-ID-B, the crystal was placed in a nitrogen cold stream on the goniometer head of a Bruker D8 diffractometer, which is equipped with a Pilatus3 X 2M detector operating in shutterless mode. Diffraction and mask data were collected using synchrotron radiation monochromate with a wavelength of 0.41328 Å with silicon (111).

Unit cell determination, integration, and scaling are then carried out in Bruker APEX 3. The precession map is integrated up to a resolution of 1.5 Å with a thickness of 0.1. Crystal structure determination and refinement were done in ShelXle[27] where an electron density map was also extracted[37]. All possible space groups and pseudo-merohedral twins are tested to find the best matching crystal structure.

*Raman Spectroscopy*: Polarization-resolved Raman spectroscopy was performed in a Renishaw inVia confocal Raman Microscope with a linearly polarized 532 nm laser and a ×50 microscope objective at room temperature. A rotating wave plate controlled the polarization of the incident beam in the incident beam path. No polarizing optics were used in the detection beam path.

Laser power density was maintained under 5 MW/m$^2$ to avoid surface heating which induces obvious degradation of the crystal. The unpolarized Raman spectrum is integrated from all polarization-resolved spectrums collected between 0° to 350°.

*Fourier Transform Infrared Spectroscopy (FTIR):* Infrared spectroscopy was performed using a Fourier-transform infrared spectrometer (Bruker Vertex 70) connected to an infrared microscope (Hyperion 2000). A 15× Cassegrain microscope objective (numerical aperture = 0.4) was used for both transmission and reflection measurements at near-normal incidence. A KRS-5 wire grid polarizer controls the polarization of the incident beam for 2-30 μm. FTIR measurements were performed with a Globar source, a potassium bromide beam splitter, and a mercury-cadmium-telluride (MCT) detector.

BaTiSe$_3$ crystals were suspended from the edge of a Si handle wafer using Kapton tape. The pre-determined crystal orientation was manually matched with the polarizer orientation. Transmittance and reflectance spectra were collected at 0° and 90° rotation from the *a*-axis.

*Spectroscopic Ellipsometry*: Variable-angle spectroscopic ellipsometry measurements were performed using a VASE ellipsometer with focusing probes (J. A. Woollam Co.) over a spectral range of 210 nm to 2500 nm at an angle of incidence of 55°. Data were acquired from four different sample orientations (optical axis perpendicular, parallel, 30°, and 18° to the plane of incidence). Data analysis and refractive index extraction were performed using WVASE software[38] (J. A. Woollam Co.). More details can be found in *Supporting Information Section V*.

*First-Principles Calculations*: Performed within the framework of density functional theory, implemented in the OpenMX package (https://www.openmx-square.org/). The pseudoatomic orbitals[39] are generated with the cutoff radius of 10 a. u., 7.0 a. u., and 7.0 a. u., for Ba, Ti, and Se elements, and basis sets of $s3p2d2$, $s3p2d1$, and $s3p2d2$ for these three elements[40], which have been tested good enough to describe our system. The exchange correlation energy functional is adopted, which is parameterized by Perdew, Burke, and Ernzerhof (PBE)[41] within the generalized gradient approximation. Internal atomic positions are relaxed starting from the experimental lattice constants. Spin-orbit coupling is included self-consistently through the *j*-dependent pseudopotential, with the $\Gamma$-centered *k*-mesh grid of 5×5×9.

**Supporting Information**

Supporting Information is available from the Wiley Online Library or from the author.

**Acknowledgments**

The authors appreciate the guidance from Dr. Gwan-Yeong Jung, and Prof. Rohan Mishra from the Washington University at St. Louis. This work was supported by the Army Research Office (ARO) MURI program with award number W911NF-21-1-0327 and the US National Science Foundation with award number DMR-2122071. The crystal growth capabilities were in part supported by an ONR grant with award number N00014-23-1-2818. H.M. and M.K. acknowledge the support from the Office of Naval Research (N00014-20-1-2297). The preliminary crystal structure is screened by the single crystal diffraction instrumentation supported by National Science Foundation award number CHE-2018740 and by the Raman spectroscopy supported by the U.S. Department of Energy, Office of Basic Energy Sciences under Award No. DE-FG02-0746376. This research used resources from the Advanced Light Source, which is a DOE Office of Science User Facility under contract no. DE-AC02-05CH11231; and the NSF's ChemMatCARS Sector 15 at the Advanced Photon Source (APS),

Argonne National Laboratory (ANL), which is supported by the Divisions of Chemistry (CHE) and Materials Research (DMR), National Science Foundation, under grant number NSF/CHE-1834750. The authors gratefully acknowledge the use of facilities and instrumentation at the UW-Madison Wisconsin Centers for Nanoscale Technology (wcnt.wisc.edu) partially supported by the NSF through the University of Wisconsin Materials Research Science and Engineering Center (DMR-1720415). The authors gratefully acknowledge the use of facilities at the Core Center for Excellence in Nano Imaging at the University of Southern California for the results reported in this manuscript. B.Z acknowledges technical assistance from Mythili Surendran and Harish Kumarasubramanian in collaboration with the related projects.</seg>

**Conflict of Interest**

The authors declare no conflict of interest.

**Data Availability Statement**

The data that support the findings of this study are available from the corresponding author upon reasonable request.

Infrared optical anisotropy, *i.e.,* birefringence and dichroism, of the quasi-1D perovskite chalcogenide $A_{1+x}TiX_3$ (A = Sr, Ba; X = S, Se) is further expanded by achieving high-quality single crystals of $BaTiSe_3$. With a crystal structure similar to $BaTiS_3$ ($\Delta n$ ~0.76), $BaTiSe_3$ possesses even larger optical anisotropy, whose dichroism ($\Delta \kappa$) peaks within SWIR and birefringence ($\Delta n$) is as large as ~0.9 in the MWIR to LWIR spectra range.

Boyang Zhao, Hongyan Mei, Zhengyu Du, Shantanu Singh, Tieyan Chang, Nick Settineri, Simon J. Teat, Yu-Sheng Chen, Stephen B. Cronin, Mikhail A. Kats, and Jayakanth Ravichandran*

Infrared Optical Anisotropy in Quasi-1D Hexagonal Chalcogenide $BaTiSe_3$

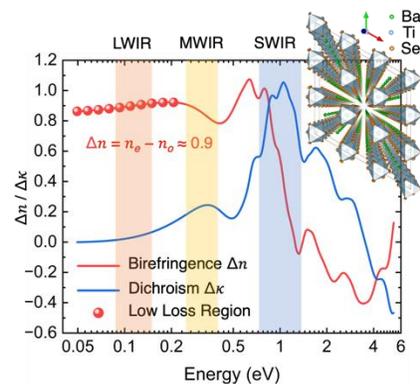

ToC figure ((Please choose one size: 55 mm broad × 50 mm high **or** 110 mm broad × 20 mm high. Please do not use any other dimensions))